\documentclass[12pt]{article}
\usepackage{amsmath}
\usepackage{latexsym}
\usepackage{amssymb}
\usepackage{a4wide}
\usepackage{bbm}
\usepackage{epsfig}
\newcommand{\I}{\text{i}}
\newcommand{\E}{\text{e}}

\newcommand{\Tr}{\text{Tr}}
\newcommand{\re}[1]{~(\ref{#1})}\begin{document}
\newcommand{\meff}{m_{\text{eff}}}
\newcommand{\Geff}{\Gamma_{\text{eff}}}
\newcommand{\Leff}{{\cal L}_{\text{eff}}}
\newcommand{\mn}{\mathbf{n}}
\newcommand{\mW}{\mathbf{W}}
\newcommand{\mA}{\mathbf{A}}
\newcommand{\mQ}{\mathbf{Q}}
\newcommand{\mK}{\mathbf{K}^{\!\mW}}
\newcommand{\mX}{\boldsymbol{\chi}}
\newcommand{\mF}{\boldsymbol{\phi}}
\newcommand{\case}[2]{{\scriptstyle \frac{#1}{#2}}}
\unitlength=1mm
\title{\bf Wilsonian effective action for SU(2) Yang-Mills theory with
  Cho-Faddeev-Niemi-Shabanov decomposition} 
\author{Holger Gies\thanks{Emmy Noether fellow}\\
  \small\it Institut f\"ur theoretische Physik, Universit\"at T\"ubingen,
  D-72076 T\"ubingen, Germany\\
  \small\it and\\
  \small\it Theory Division, CERN, CH-1211 Geneva, Switzerland\\
  \small\it E-mail: holger.gies@cern.ch }
\maketitle
\begin{abstract}
  The Cho-Faddeev-Niemi-Shabanov decomposition of the SU(2) Yang-Mills
  field is employed for the calculation of the corresponding Wilsonian
  effective action to one-loop order with covariant gauge fixing. The
  generation of a mass scale is observed, and the flow of the marginal
  couplings is studied. Our results indicate that higher-derivative
  terms of the color-unit-vector $\mathbf{n}$ field are necessary for
  the description of topologically stable knotlike solitons which
  have been conjectured to be the large-distance degrees of freedom.
\end{abstract}

\section{Introduction}
The fact that quarks and gluons are not observed as asymptotic states
in our world indicates that a description in terms of these fields
is not the most appropriate language for discussing
low-energy QCD. On the other hand, there seems to be little predictive
virtue in describing the low-energy domain only by observable
quantities, such as mesons and baryons. A purposive procedure can be
the identification of those (not necessarily observable) degrees of
freedom of the system that allow for a ``simple'' description of the
observable states. The required ``simplicity'' can be measured in
terms of the simplicity of the action that governs those degrees of
freedom. Clearly, a clever guess of such degrees of freedom is halfway
to the solution of the theory; the remaining problem is to prove that
these degrees of freedom truly arise from the fundamental theory by
integrating out the high-energy modes.

For the pure Yang-Mills (YM) sector of QCD, such a guess has recently
been made by Faddeev and Niemi \cite{Faddeev:1999eq} inspired by the
work of Cho \cite{Cho:1980nv}. For the gauge group SU(2), they
decomposed the (implicitly gauge-fixed) gauge potential
$\mA_\mu$ into an ``abelian'' component $C_\mu$, a unit color vector
$\mn$ and a complex scalar field $\varphi$; here, $C_\mu$ is the local
projection of $\mA_\mu$ onto some direction in color space defined by
the space-dependent $\mn$. Faddeev and Niemi conjectured that the
important low-energy dynamics of SU(2) YM theory\footnote{Different
  generalizations of the gauge field decomposition for higher gauge
  groups can be found in \cite{Periwal:1998pc}, \cite{Faddeev:1999yz}
  and \cite{Shabanov:1999xy}.} is determined by the
$\mn$ field; its effective action of nonlinear sigma-model type,
the Skyrme-Faddeev model, should then arise from integrating out the
further degrees of freedom: $C_\mu$, $\varphi$, \dots:
\begin{equation}
\Geff^{\text{FN}}= \int d^4x\, \left[ m^2 (\partial_\mu \mn)^2 +
  \frac{1}{g^2} (\mn\cdot
  \partial_\mu\mn\times\partial_\nu\mn)^2\right]. \label{1}
\end{equation}
The additional mass scale $m$ is expected to be generated by the
integration process as well; first hints of this mechanism have been
observed in a one-loop integration over a reduced set of variables
\cite{Langmann:1999nn,Cho:1999wp}. The associated knotlike solitonic
excitations of the Skyrme-Faddeev model are supposed to be identified
with glue balls (which are directly observable at least on the
lattice).\footnote{In a very recent paper \cite{Faddeev:2001dd},
  Faddeev and Niemi generalized their decomposition in order to obtain
  a manifest duality between the here-considered ``magnetic'' and
  additional ``electric'' variables, involving an abelian scalar
  multiplet with two complex scalars. This electric sector will not be
  considered in the present work.}

The presence of gauge symmetry in YM theory complicates this ambitious
conjecture in two ways: first, in order to formulate a quantum theory,
the decomposition of $\mA_\mu$ has to also include the overabundant
gauge degrees of freedom; and secondly, the gauge has then to be fixed
in a prescribed way, not only to be able to perform functional
integration, but also to arrive nevertheless at a unique $\mn$
field.\footnote{A different approach was put forward in
  \cite{Khvedelidze:1999hm}, where the $\mn$ field was identified by
  constructing an unconstraint version of SU(2) Yang-Mills theory in a
  Hamiltonian context.}

The first problem was solved by Shabanov
\cite{Shabanov:1999xy,Shabanov:1999uv}, who established a one-to-one
correspondence between the unfixed gauge field $\mA_\mu$ and its
decomposition, and the quantum theory was formulated; his results are
briefly sketched in Sect. 2 and shall serve as the starting point of
our investigations. The second problem of gauge fixing implies that a
successful realization of the ideas of Faddeev and Niemi will only be
meaningful in a certain gauge. In this (a priori unknown) gauge, the
important low-energy degrees of freedom might in fact be determined by
the $\mn$ field and a simple action, whereas in a different gauge,
these degrees of freedom may be hidden in a highly complicated
structure involving the $\mn$ and other fields.

The present paper is dedicated to a calculation of the one-loop
Wilsonian effective action for SU(2) Yang-Mills theory in terms of the
gauge field decomposition of Shabanov. Our intention is to study the
renormalization group flow of the mass scale parameter of Eq.\re{1},
the gauge coupling and further marginal couplings. In view of the
second problem mentioned above, our results and their interpretation
are strictly tied to the particular gauge we shall choose. We
face this problem by fixing the gauge in such a way that Lorentz
invariance and global color transformations remain as residual
symmetries; these are the symmetries of the Skyrme-Faddeev model and
must mandatorily be respected.

The Wilsonian effective action is characterized by the fact that it
governs the dynamics of the low-energy modes below a certain cutoff
$k$; it incorporates the interactions that are induced by high-energy
fluctuations with momenta between $k$ and the ultraviolet (UV) cutoff
$\Lambda$ which have been integrated out. Following the Faddeev-Niemi
conjecture, we only retain the $\mn$ field as low-energy degree of
freedom. Actually, we integrate over the high-energy modes in two
different ways: first, we integrate out the $k<p<\Lambda$ fluctuations
of all fields {\em except for} the $\mn$ field, which is left
untouched (Sec.~3). Secondly, we integrate all fields {\em including}
the $\mn$ field over the same momentum shell (Sec.~4). In this way, we
can study the effect of the $\mn$ field fluctuations on the flow of
the mass scale and the couplings in detail.

The results for both calculations are similar: the mass scale $m$
appearing in Eq.\re{1} is indeed generated by the renormalization
group flow, and the gauge coupling is asymptotically free. As far as
the simplicity of the conjectured effective action Eq.\re{1} is
concerned, our results are a bit disappointing: as discussed in
Sec.~5, further marginal terms (not displayed in Eq.\re{1}) are of the
same order as the displayed one and therefore have to be included in
Eq.\re{1}. Keeping only those terms that involve single
derivatives acting on $\mn$ results in an action without stable
solitons; nevertheless, stability is in fact ensured owing to the
presence of higher-derivative terms. The disadvantage is that these
terms spoil the desired simplicity of the low-energy effective
theory. 

Of course, our perturbative results represent only a first glance at
the true infrared behavior of the system and are far from providing
qualitatively confirmed results, not to mention quantitative
predictions. To be precise, the one-loop calculation investigates only
the form of the renormalization group trajectories of the couplings in
the vicinity of the perturbative Gaussian fixed point. Nevertheless,
various extrapolations of the perturbative trajectories can elucidate
the question as to whether the Faddeev-Niemi conjecture is realizable
or not.

\section{Quantum Yang-Mills theory in Cho-Faddeev-Niemi-Shabanov
  variables} 

In decomposing the Yang-Mills gauge connection, we follow
\cite{Cho:1980nv,Shabanov:1999xy,Shabanov:1999uv}. Let $\mA_\mu$ be an
SU(2) connection where the color degrees of freedom are represented in
vector notation. We parametrize $\mA_\mu$ as
\begin{equation}
\mA_\mu=\mn\, C_\mu +(\partial_\mu\mn)\times\mn +\mW_\mu,\label{2}
\end{equation}
where the cross product is defined via the SU(2) structure constants.
$C_\mu$ is an ``abelian'' connection, whereas $\mn$ denotes a unit
vector in color space, $\mn\cdot\mn=1$. $\mW_\mu$ shall be orthogonal
to $\mn$ in color space, obeying $\mW_\mu\cdot\mn=0$, so that
$C_\mu=\mn\cdot\mA_\mu$. For a given $\mn$, $C_\mu$ and $\mW_\mu$, the
connection $\mA_\mu$ is uniquely determined by Eq.\re{2}. In the
opposite direction, there is still some arbitrariness: for a given
$\mA_\mu$, $\mn$ can generally be chosen at will, but then $C_\mu$ and
$\mW_\mu$ are fixed (e.g., $\mW_\mu=\mn\times D_\mu(\mA)\mn$, where
$D_\mu$ denotes the covariant derivative).

While the LHS of Eq.\re{2} describes $3_{\text{color}}\times
4_{\text{Lorentz}}=12$ off-shell and gauge-unfixed degrees of freedom,
the RHS up to now allows for $(C_\mu\!:)4_{\text{Lorentz}}
+(\mn\!:)2_{\text{color}} +(\mW_\mu\!:) 3_{\text{color}}\times
4_{\text{Lorentz}}-4_{\mn\cdot\mW_\mu=0} =14$ degrees of freedom. Two
degrees of freedom on the RHS remain to be fixed. For example, by
fixing $\mn$ to point along a certain direction and imposing gauge
conditions on $\mW_\mu$, we arrive at the class of abelian gauges
which are known to induce monopole degrees of freedom in $C_\mu$. In
order to avoid these topological defects, we let $\mn$ vary in
spacetime and impose a general condition on $C_\mu,\mn$ and $\mW_\mu$,
\begin{equation}
\mX(\mn,C_\mu,\mW_\mu)=0, \qquad\text{with}\quad
\mX\cdot\mn=0, \label{3}
\end{equation}
which fixes the redundant two degrees of freedom on the RHS of
Eq.\re{2}. Moreover, Eq.\re{3} also determines how $\mn$, $C_\mu$ and
$\mW_\mu$ transform under gauge transformations of $\mA_\mu$: by 
demanding that $\delta\mX(\mn,C_\mu(\mA),\mW_\mu(\mA))=0$ (and
$\delta(\mX\cdot\mn)=0$), the transformation $\delta \mn$ of $\mn$ is
uniquely determined, from which $\delta C_\mu$ and $\delta\mW_\mu$ are
also obtainable.

The thus established one-to-one correspondence between $\mA_\mu$ and
its decomposition\re{2} allows us to rewrite the generating functional
of YM theory in terms of a functional integral over the new fields
\cite{Shabanov:1999xy,Shabanov:1999uv}:
\begin{equation}
Z=\int{\cal D}\mn{\cal D}C{\cal D}\mW\, \delta(\mX)\,
\Delta_{\text{S}}\, \Delta_{\text{FP}}\,\E^{-S_{\text{YM}}
  -S_{\text{gf}}}.\label{5} 
\end{equation}
Beyond the usual Faddeev-Popov determinant $\Delta_{\text{FP}}$, the
YM action $S_{\text{YM}}$ and the gauge fixing action $S_{\text{gf}}$,
we find one further determinant introduced by Shabanov, 
$\Delta_{\text{S}}$; this determinant accompanies the $\delta$
functional which enforces the constraint $\mX=0$, in complete analogy
to the Faddeev-Popov procedure:
\begin{equation}
\Delta_{\text{S}}:= \text{det}\, \left(\left.\frac{\delta \mX}{\delta
      \mn} \right|_{\mX=0} \right). \label{6}
\end{equation}
All objects in the integrand of Eq.\re{5} are understood to be
functions of the 14 integration variables $\mn$, $C_\mu$ and
$\mW_\mu$.

By construction, the generating functional\re{5} is invariant under
different choices of $\mX$ for the same reason that it is invariant
under different choices of the gauge -- this is controlled by the
Faddeev-Popov procedure.

Nevertheless, the choice of $\mX$ crucially belongs to the definition
of the decomposition\re{2} and of the conjectured low-energy degrees
of freedom; in other words, even if there is one particular $\mX$ that
leads to Eq.\re{1} as the true low-energy effective action after
integrating out $C_\mu$ and $\mW_\mu$, other choices of $\mX$ will not
lead to the same result, because the low-energy degrees of freedom
then are differently distributed over $\mn$, $C_\mu$ and $\mW_\mu$.

In the present work, $\mX$ is chosen in such a way that
$\mn$ transforms homogeneously under gauge transformations, i.e.,
$\mn$ is orthogonally rotated in color space \cite{Cho:1980nv}:
\begin{eqnarray}
0&=&\mX:= \partial_\mu\mW_\mu+ C_\mu\mn\times\mW_\mu +\mn (\mW_\mu\cdot
\partial_\mu\mn), \label{4}\\
\Rightarrow&&\delta\mn=\mn\times \boldsymbol{\varphi},
\quad\text{under}\,\,\delta
\mA_\mu=D_\mu(\mA)\boldsymbol{\varphi}=\partial_\mu \boldsymbol{\varphi}
+\mA_\mu\times\boldsymbol{\varphi}. \nonumber
\end{eqnarray}
Incidentally, the gauge transformation properties of $C_\mu$ and
$\mW_\mu$ also become very simple with the choice\re{4}: $\mW_\mu$
also transforms homogeneously, and $\delta C_\mu=\mn\cdot\partial_\mu
\boldsymbol{\varphi}$. 

Finally, the choice of the gauge-fixing condition must also be viewed
as being part of the definition of the decomposition. Not only does
the functional form of $\Delta_{\text{FP}}$ and $S_{\text{gf}}$
depend on this choice, but the discrimination of high- and
low-momentum modes is also determined by the gauge fixing. In fact,
this gauge dependence of the mode momenta usually is the main obstacle
against setting up a Wilsonian renormalization group study. But in the
present context, it belongs to the conjecture that the particular
gauge that we shall choose singles out those low-momentum modes which
finally provide for a simple description of low-energy QCD; in a
different gauge, we would encounter different low-momentum modes, but
we also would not expect to find the same simple description. 

In this work, we choose the covariant gauge condition $\partial_\mu
\mA_\mu=0$. This automatically ensures covariance of the resulting
effective action and, moreover, allows for the residual symmetry of
global gauge transformations, $\boldsymbol{\varphi}=\,\,$const.
Together with the choice\re{4}, this residual symmetry coincides with
the desired global color symmetry of the Skyrme-Faddeev model\re{1}.
This means that the demand for color and Lorentz symmetry of the
action\re{1} is satisfied exactly by a covariant gauge and
condition\re{4}.

\section{One-loop effective action without $\mn$ fluctuations}
\label{without}

Our aim is the construction of the one-loop Wilsonian effective action
for the $\mn$ field by integrating out the $C$ and $\mW$ field over a
momentum shell between the UV cutoff $\Lambda$ and an infrared cutoff
$k<\Lambda$. In general, this will induce nonlinear and nonlocal
self-interactions of the $\mn$ field; since we are looking for an
action of the type\re{1}, we represent these interactions in a
derivative expansion and neglect higher derivative terms of order
${\cal O}(\partial^2 \mn \partial^2 \mn)$ (later, we shall question
this approach).

Furthermore, we do not integrate out $\mn$ field fluctuations in this
section (see Sect.~\ref{nfluc}) and disregard any induced $C$ or $\mW$
interactions below the infrared cutoff $k$. From a technical
viewpoint, the one-loop approximation of the desired effective action
$\Gamma_k[\mn]$ is obtained by a Gaussian integration of the quadratic
$C$ and $\mW$ terms in Eq.\re{5}, neglecting higher-order terms of the
action:
\begin{eqnarray}
\E^{-\hat{\Gamma}_k[\mn]} &=&\E^{-S_{\text{cl}}[\mn]} 
\int_k {\cal  D}C{\cal D}\mW\, \Delta_{\text{S}}[\mn]\,
\Delta_{\text{FP}}[\mn] \delta(\mX)  \label{7}\\
&&\qquad\qquad \times
  \E^{-\frac{1}{g^2}\int \left\{ C_\mu \case{1}{2} M^C_{\mu\nu} C_\nu
  +\mW_\mu \case{1}{2} M^{\mW}_{\mu\nu}\mW_\nu +C_\mu
  \mQ^C_{\mu\nu}\cdot \mW_\nu +C_\nu K^C_\nu +\mW_\mu \cdot
  \mK_\mu\right\} }, \nonumber 
\end{eqnarray}
where the hat on $\hat{\Gamma}_k[\mn]$ indicates that the $\mn$ field
fluctuations have not been taken into account. Furthermore, any $C$ or
$\mW$ dependence of $\Delta_{\text{S}}$ and $\Delta_{\text{FP}}$ has
been neglected to one-loop order; the various differential operators
and currents which all depend on $\mn$ (and the gauge parameter
$\alpha$) are defined in Appendix A. The classical action of $\mn$
including gauge fixing terms is given by:
\begin{equation}
S_{\text{cl}}[\mn]:= \int d^4x\, \left( \frac{1}{4g^2}
  (\partial_\mu\mn \times \partial_\nu\mn)^2 + \frac{1}{2\alpha g^2}
  (\partial^2\mn \times \mn)^2\right). \label{8}
\end{equation}
We treat the $\delta$ functional in Eq.\re{7} in its Fourier
representation,
\begin{equation}
\delta(\mX)\to \int{\cal D}\mF \,\E^{-\I \int \mF\cdot\partial_\mu
  \mW_\mu +\mF\cdot C_\mu \mn\times \mW_\mu+ (\mF\cdot\mn)(\partial_\mu \mn
  \cdot \mW_\mu)}, \label{9}
\end{equation}
where the second term in the exponent, the triple vertex, can actually
be neglected, because it leads only to nonlocal terms (cf. later) or
terms of higher order in derivatives. Inserting Eq.\re{9} into
Eq.\re{7}, we end up with three functional integrals over $C$, $\mW$
and $\mF$, which can successively be performed, leading to three
determinants,
\begin{equation}
\E^{-\hat{\Gamma}_k[\mn]} \to\E^{-S_{\text{cl}}[\mn]}
\Delta_{\text{S}}[\mn]\,\Delta_{\text{FP}}[\mn] \, \bigl( \det
M^C\bigr)^{-1/2} \, \bigl(\det \overline{M}^{\mW}\bigr)^{-1/2}\, \bigl(
\det - \widetilde{Q}^{\mF}_\mu (\overline{M}^{\mW})^{-1}_{\mu\nu}
Q^{\mF}_\nu \bigr)^{-1/2}, \label{10}
\end{equation}
where we have omitted several nonlocal terms that arise from the
completion of the square in the exponent during the Gaussian
integration. In Appendix B, we argue that these nonlocal terms are
unimportant in the present Wilsonian investigation. Again, details
about the various operators in Eq.\re{10} are given in App. A.

The determinants are functionals of $\mn$ only and have to be
evaluated over the space of test functions with momenta between $k$
and $\Lambda$. The determinants depend also on the gauge parameter
$\alpha$. Only for the Landau gauge $\alpha=0$ is the gauge-fixing
$\delta$ functional implemented exactly; in fact, $\alpha=0$ appears
to be a fixed point of the renormalization group flow
\cite{Ellwanger:1996qf}. But this in turn ensures that the choice of
$\alpha=\alpha(k)\equiv\alpha_k$ at the cutoff scale $k\to\Lambda$ is
to some extent arbitrary, since $\alpha_k$ flows to zero anyway as $k$
is lowered. This allows us to conveniently choose
$\alpha_{k=\Lambda}=1$ at the cutoff scale and evaluate the
determinants with this parameter choice.

As mentioned above, we evaluate the determinants in a derivative
expansion based on the assumption that the low-order derivatives of
$\mn$ represent the essential degrees of freedom in the low-energy
domain. There are various techniques for the calculation at our
disposal; it turns out that a direct momentum expansion of the
operators is most efficient.\footnote{As cross-checks, we also
  employed a propertime representation for the operators which we
  decomposed with a heat-kernel expansion as well as with a multiple
  use of the Baker-Campbell-Hausdorff formula.} We shall demonstrate
this method by means of the third determinant of Eq.\re{10}, the ``$C$
determinant''; the key observation is that derivatives acting on the
space of test functions create momenta of the order of $p$ with
$k<p<\Lambda$, whereas derivatives of the $\mn$ field are assumed to
obey $|\partial\mn| \ll k$ in agreement with the Faddeev-Niemi
conjecture. This suggests an expansion of the form
\begin{eqnarray}
\ln \bigl(\det M^C\bigr)^{1/2}&=& -\frac{1}{2} \Tr\, \ln \bigl(
-\partial^2 \mathbbm{1}_{\text{L}} +\partial\mn \cdot
\partial\mn\bigr)\nonumber\\
&=&-\frac{1}{2} \Tr\, \left[ \ln (-\partial^2\mathbbm{1}_{\text{L}})
  +\ln \left( \mathbbm{1}_{\text{L}} + \frac{\partial\mn\cdot
      \partial\mn}{-\partial^2} \right) \right]\label{11}\\
&=& -\frac{1}{2} \Tr\, \ln (-\partial^2\mathbbm{1}_{\text{L}})
-\frac{1}{2} \Tr\, \frac{\partial\mn\cdot \partial\mn}{-\partial^2}
+\frac{1}{4} \Tr\, \left(\frac{\partial\mn\cdot
    \partial\mn}{-\partial^2} \right)^2 + {\cal O}((\partial\mn)^6),
 \nonumber
\end{eqnarray}
where we suppressed Lorentz (L) indices. Here, we neglected
higher-derivative terms of $\mn$, e.g., $\partial^2\mn$, which is in
the spirit of the Faddeev-Niemi conjecture; of course, this has to be
checked later on. Employing the integral formulas given in App. C, we
finally obtain for the $C$ determinant 
\begin{eqnarray}
\ln\bigl(\det M^C\bigr)^{1/2} &\simeq &-\frac{1}{32 \pi^2}
(\Lambda^2-k^2) \int_x (\partial_\mu\mn)^2 \nonumber\\
&&-\frac{1}{32\pi^2}\ln\frac{\Lambda}{k} \int_x \bigl(
\partial_\mu\mn\times\partial_\nu\mn\bigr)^2 
+ \frac{1}{32\pi^2} \ln\frac{\Lambda}{k} \int_x (\partial_\mu\mn)^4,
\label{12}
\end{eqnarray}
where $\int_x\equiv \int d^4x$. The first term contributes to the
desired mass term of Eq.\re{1}, whereas the second and third
renormalize the classical action\re{8}. 

The remaining four determinants of Eq.\re{10} have to be evaluated in
the same way. The calculation is straightforward though
extensive. Here, we shall cite only the final results:
\begin{eqnarray}
\ln \Delta_{\text{FP}}\!\!&=&\!\! 
-\frac{(\Lambda^2\!-\!k^2)}{64\pi^2} \int_x \!(\partial_\mu\mn)^2 
+ \frac{1}{48\pi^2} \ln \frac{\Lambda}{k}\int_x\!  
  \bigl(\partial_\mu\mn\!\times\!\partial_\nu\mn\bigr)^2 
- \frac{1}{32\pi^2} \ln\frac{\Lambda}{k} \int_x \!(\partial_\mu\mn)^4,
\nonumber\\
\ln (\det\overline{M}^{\mW}\!)^{-1/2}\!\!\!\!
&=&\!\!\! -\frac{5 (\Lambda^2\!-\!k^2)}{64\pi^2}\! \int_x\!
  (\partial_\mu\mn)^2 \!
- \frac{5}{24\pi^2} \ln \frac{\Lambda}{k} \int_x \!
\bigl(\partial_\mu\mn\!\times\!\partial_\nu\mn\bigr)^2 \!
+ \frac{35}{128\pi^2} \ln\frac{\Lambda}{k} \int_x
  \!(\partial_\mu\mn)^4\!, 
\nonumber\\
\ln(\det-\widetilde{Q}^{\mF}\overline{M}^{\mW}
\!\!\!\!\!&{}^{{}^{\scriptstyle -1}}&\!\!\!\!\!Q^{\mF})^{-1/2} 
= \frac{3  (\Lambda^2\!\!-\!k^2)}{128\pi^2}\! \int_x\! 
  (\partial_\mu\mn)^2 \!
+ \frac{49}{192\pi^2} \ln \frac{\Lambda}{k} \int_x \!
\bigl(\partial_\mu\mn\!\times\!\partial_\nu\mn\bigr)^2 \nonumber\\
&&\qquad\qquad\,\,- \frac{5}{16\pi^2} \ln\frac{\Lambda}{k} \int_x
  \!(\partial_\mu\mn)^4\!. 
\label{13}
\end{eqnarray}
The determinant $\Delta_{\text{S}}$ does not contribute, because it is
independent of $\mn$ in one-loop approximation. Inserting these
results into Eq.\re{10} leads us to the desired Wilsonian effective
action to one-loop order for the $\mn$ field in a derivative
expansion:
\begin{eqnarray}
\hat{\Gamma}_k[\mn] &=& \frac{13}{8} \frac{\Lambda^2}{16\pi^2} \bigl(
1-\E^{2t}\bigr) \int_x (\partial_\mu \mn)^2 + \frac{1}{4} \left(
  \frac{1}{g^2} + \frac{7}{3} \frac{1}{16\pi^2} \,t\right)
\int_x(\partial_\mu\mn \times \partial_\nu \mn)^2 \nonumber\\
&&-\frac{1}{2} \left( \frac{1}{\alpha g^2} + \frac{5}{4}
  \frac{1}{16\pi^2} \, t\right) \int_x (\partial_\mu \mn)^4,
\label{14}
\end{eqnarray}
where $t=\ln k/\Lambda\in ]-\infty,0]$ denotes the ``renormalization
group time''. We would like to stress once more that
$\hat{\Gamma}_k[\mn]$ does not contain the result of fluctuations of
the $\mn$ field itself; in other words, it represents (an
approximation to) the ``tree-level action'' for the complete quantum
theory of the $\mn$ field. 

Indeed, the generation of a ``kinetic'' term $\sim
(\partial_\mu\mn)^2$ growing under the flow of increasing $k$ as
conjectured by Faddeev and Niemi is observed. Moreover, it has the
correct sign (+), implying that an ``effective field theory''
interpretation seems possible. The second term which is proportional
to the classical action reveals information about the renormalization
of the Yang-Mills coupling:
\begin{equation}
\frac{1}{\hat{g}_{\text{R}}^2} := \frac{1}{g^2} + \frac{7}{3}
\frac{1}{16\pi^2} \, t\quad \Rightarrow\quad
\hat{\beta}_{g^2}:=\partial_t \hat{g}_{\text{R}}^2 =-\frac{7}{3}
\frac{1}{16\pi^2} \, \hat{g}_{\text{R}}^4. \label{15}
\end{equation}
The resulting $\hat{\beta}$ function is a factor of $44/7$ smaller
than the $\beta$ function of full Yang-Mills theory for SU(2). This is
an expected result, since we did not integrate over all degrees of
freedom of the gauge field; the $\mn$ integration still remains.
Nevertheless, the $\hat{\beta}$ function implies asymptotic freedom, 
which indicates that the decomposition of the Yang-Mills field is not
a pathologically absurd choice. It is interesting to observe that the
$C$ and $\mW$ determinants contribute positively to
$\hat{\beta}_{g^2}$, whereas the Faddeev-Popov and the $\mF$
determinant contribute negatively; the latter, which arises from the
$\mW$ fixing, even dominates: $-7/3=[6_C-4_{\text{FP}} +40_{\mW}
-49_{\mF}]/3$.

The third term of Eq.\re{14} contains information about the
renormalization of the gauge parameter $\alpha$ under the flow:
\begin{equation}
\frac{1}{\hat{\alpha}_{\text{R}} \hat{g}_{\text{R}}^2}
=\frac{1}{\alpha g^2} + \frac{5}{4} \frac{1}{16\pi^2}\,
t,\quad\Rightarrow\quad 
\partial_t \hat{\alpha}_{\text{R}} = \frac{7}{3}
\hat{\alpha}_{\text{R}} \left( 1- \frac{15}{28}
  \hat{\alpha}_{\text{R}}\right)
\frac{\hat{g}_{\text{R}}^2}{16\pi^2}. \label{16} 
\end{equation}
The RHS of this renormalization group equation is positive for
$\alpha<{28/15}\simeq1.87$; this implies that $\alpha$ runs
to zero under the flow as long as
$\alpha_\Lambda<{28/15}$. Therefore, our starting point
$\alpha_\Lambda=1$ is a consistent choice that ensures a running into
the desired Landau gauge $\alpha\to 0$. 

Before we discuss the physical implications of our result Eq.\re{14},
let us study the effective action including the $\mn$ field
fluctuations. In principle, this action should be obtainable from the
present result by inserting $\hat{\Gamma}_k[\mn]$ into a functional
integral over $\mn$. However, we evaluated $\hat{\Gamma}_k[\mn]$ in a
derivative expansion, neglecting high-momentum fluctuations of the
$\mn$ field. But when integrating over $\mn$ fluctuations, especially
these high-momentum modes are important for the renormalization of the
couplings. Hence, their correct running cannot be calculated via such
an indirect approach. The direct way is presented in the next section.

\section{One-loop effective action including $\mn$ fluctuations}
\label{nfluc}

In the following, we propose a different way to integrate out the
``hard'' modes with high momenta $p$, $k<p<\Lambda$. This time, we
also include the hard fluctuations of the $\mn$ field and decompose
the complete Yang-Mills field into soft and hard modes,
\begin{equation}
\mA_\mu=\mA^{\text{S}}_\mu +\mA^{\text{H}}_\mu, \quad 
\mA^{\text{S,H}}_\mu=\mA^{\text{S,H}}_\mu 
(C^{\text{S,H}}_\mu,\mn^{\text{S,H}},\mW^{\text{S,H}}_\mu). \label{17}
\end{equation}
Since the hard modes $\mA^{\text{H}}_\mu$ shall be integrated out
completely, the explicit use of the decomposition into
$C^{\text{H}}_\mu$, $\mn^{\text{H}}$ and $\mW^{\text{H}}_\mu$ would be
a very inconvenient choice of overabundant integration
variables; therefore, the decomposition is only adopted for the soft
modes $\mA^{\text{S}}_\mu$. In the spirit of the Faddeev-Niemi
conjecture, we assume that these soft modes are dominated by the $\mn$
field:
\begin{equation}
\mA^{\text{S}}_\mu =\partial_\mu\mn^{\text{S}}\times
\mn^{\text{S}}. \label{18}
\end{equation}
Integrating out the hard modes $\mA^{\text{H}}$ results in two
determinants in one-loop approximation,
\begin{equation}
\Gamma_k[\mA^{\text{S}}] =\frac{1}{2} \ln \det
(\Delta^{\text{YM}}[\mA^{\text{S}}])^{-1} 
-\ln\det\Delta_{\text{FP}}[A^{\text{S}}], \label{19}
\end{equation}
corresponding to the hard gluon and ghost loops; again we dropped the
nonlocal terms (cf.~App.~B). The ghost contribution in the form of the
Faddeev-Popov determinant is, of course, identical to the one obtained
in the first line of Eq.\re{13}, since the gauge fixing is performed
in the same way as before. The gluonic determinant involves the
operator
\begin{equation}
(\Delta^{\text{YM}}[\mA^{\text{S}}])^{-1}_{\mu\nu} =-\left[ D^2\,
  \mathbbm{1}_{\text{L}}-2\I F-D D+ \frac{1}{\alpha} \partial \partial
  \right]_{\mu\nu} \bigg|_{\mA=\mA^{\text{S}}}, \label{20}
\end{equation} 
where $D_\mu$ denotes the covariant derivative and $F_{\mu\nu}$ the
field strength tensor. The explicit representation of Eq.\re{20} in
terms of the $\mn$ field is again given in App.~A, Eqs.\re{A5}
and\re{A6}. The determinants in Eq.\re{19} can be calculated in a
derivative expansion in the same way as described in the preceding
section. Since the computation of the term $\sim (\partial\mn)^2$ is
already very laborious, we do not calculate the marginal terms $\sim
(\partial_\mu\mn\times\partial_\nu\mn)^2$ etc. directly, but take over
the known one-loop results for the running coupling and the gauge
parameter from \cite{Ellwanger:1996qf}. The final result for the
Wilsonian one-loop effective action for the soft modes of the $\mn$
field reads
\begin{eqnarray}
{\Gamma}_k[\mn] &=& \frac{\Lambda^2}{16\pi^2} \bigl(
  1-\E^{2t}\bigr) \int_x (\partial_\mu \mn)^2 
+ \frac{1}{4} \left( \frac{1}{g^2} + \frac{44}{3} \frac{1}{16\pi^2}
  \,t\right) \int_x(\partial_\mu\mn \times \partial_\nu \mn)^2
  \label{21}\\ 
&&-\frac{1}{2} \left( \frac{1}{\alpha g^2} + \frac{14}{3}
  \frac{1}{16\pi^2} \, t\right) \int_x (\partial_\mu \mn)^4
+\frac{1}{2} \left( \frac{1}{\alpha g^2} + \frac{14}{3}
  \frac{1}{16\pi^2} \, t\right)
  \int_x(\partial^2\mn\cdot\partial^2\mn), \nonumber
\end{eqnarray}
where we dropped the superscript S. Furthermore, we included for later
use a higher-derivative term $\sim\partial^2\mn\cdot\partial^2\mn$
which is also marginal in the renormalization group sense and
accompanied by the $1/(\alpha g^2)$ coefficient in the classical action.

Again, the generation of the ``kinetic'' term $\sim(\partial\mn)^2$
with a mass scale is observed; it is smaller by a factor of $8/13$ 
than in the preceding section. This means that the hard $\mn$ field
fluctuations that have been taken into account in Eq.\re{21} reduce
the new mass scale slightly; on the other hand, they increase the
running of the Yang-Mills coupling by contributing the missing piece
to the $\beta$ function which now obtains the correct SU(2) value,
$\beta_{g^2}=\frac{44}{3} \frac{1}{16\pi^2} g_{\text{R}}^4$. The
running of the gauge parameter $\alpha$ is also increased, but no
qualitative changes compared to Eq.\re{14} can be observed.

\section{Discussion and Conclusions}
The main results of our paper are contained in Eqs.\re{14} and\re{21},
where the Wilsonian one-loop effective actions $\hat{\Gamma}_k$ and
$\Gamma_k$ for the $\mn$ field have been given without and including
hard $\mn$ field fluctuations, respectively. We were able to
demonstrate that a ``kinetic'' term with a new mass scale for the
$\mn$ field is indeed generated perturbatively, as was conjectured by
Faddeev and Niemi. This term is relevant in the renormalization group
sense and perturbatively exhibits a quadratic dependence on the UV
cutoff $\Lambda$.

Furthermore, we studied the renormalization group flow of the marginal
couplings of the $\mn$ field self-interactions given by the Yang-Mills
coupling and the gauge parameter. These terms are responsible for the
stabilization of possible topological excitations of the $\mn$ field,
as suggested by the Skyrme-Faddeev model. In total, the difference
between $\hat{\Gamma}_k$ and $\Gamma_k$ is only of quantitative
nature: the inclusion of hard $\mn$ field fluctuations increases the
running of the marginal couplings and reduces the new mass scale;
qualitative features such as stability of possible solitons remain
untouched. 

In fact, the question of stability turns out to be delicate:
truncating our results for $\hat{\Gamma}_k$ or $\Gamma_k$ in
Eqs.\re{14} or\re{21} at the level of the original Faddeev-Niemi
proposal Eq.\re{1} (the first lines of Eqs.\re{14} and\re{21},
respectively), we find an action that allows for stable knotlike
solitons, since the coefficients of both terms are positive (as long
as we stay away from the Landau pole, which we consider as unphysical).
Taking additionally the $(\partial\mn)^4$ term of $\hat{\Gamma}_k$ or
$\Gamma_k$ into account, which is also marginal and does not contain
second-order derivatives on $\mn$, stability is lost, since the
coupling coefficient is negative in Eqs.\re{14} and\re{21}; for stable
solitons, a strictly positive coefficient would be required for this
truncation, as was shown in \cite{Gladikowski:1997mb}.

Finally dropping the demand for first-order derivatives, we can
include one further marginal term $\sim
\partial^2\mn\cdot\partial^2\mn$ as given in Eq.\re{21} for
$\Gamma_k$. With the aid of the identity
\begin{equation}
\int_x
(\partial^2\mn\times\mn)^2=\int_x[\partial^2\mn\cdot\partial^2\mn
-(\partial_\mu\mn)^4], \label{22}
\end{equation}
we find that the second line of Eq.\re{21} represents a strictly
positive contribution to the action which again stabilizes possible
solitons.\footnote{We expect a similar behavior for the action
  $\hat{\Gamma}_k$ in Eq.\re{14}, although we have not calculated
  the coefficient of the $\partial^2\mn\cdot\partial^2\mn$ term
  explicitly.} 

Of course, this game could be continued by including further
destabilizing and stabilizing higher-order terms again and again, but
such terms are irrelevant in a renormalization group sense; that means
their corresponding couplings are accompanied by inverse powers of the
UV cutoff $\Lambda$ and are thereby expected to vanish in the limit of
large cutoff. 

To summarize, our perturbative renormalization group analysis suggests
enlarging the Faddeev-Niemi proposal for the effective low-energy
action of Yang-Mills theory by taking all marginal operators of a
derivative expansion into account. The original proposal of Eq.\re{1}
was inspired by a desired Hamiltonian interpretation of the action
that demands the absence of third- or higher-order time
derivatives. But, as demonstrated, the covariant renormalization group
does not care about a Hamiltonian interpretation of the final
result. In some sense, the desired ``simplicity'' of the final result
is spoiled by the presence of higher-derivative terms; moreover, it
remains questionable as to whether the importance of the
$\partial^2\mn\cdot\partial^2\mn$ term is still consistent with the
derivative expansion of the action. Unfortunately, this cannot be
checked within the present approach.

It should be stressed once again that the perturbative investigation
performed here hardly suffices to confirm results about the infrared
domain of Yang-Mills theories. On the contrary, it is only a valid
approximation in the vicinity of the Gaussian UV fixed point of the
theory. Nevertheless, our study might lend some intuition to 
possible nonperturbative scenarios: for example, let us assume that
the Landau gauge $\alpha=0$ indeed is an infrared fixed point in
covariant gauges. Then the stabilizing term $\sim
(\partial^2\mn\times\mn)^2$ is enhanced in the infrared, provided that
the increase of the running coupling $g$ obeys $\alpha g^2\to 0$ for
$k\to 0$; this would be realized, e.g., if $g$ approached an infrared
fixed point. Such a scenario thus supports the idea of topological
knotlike solitons as important infrared degrees of freedom of
Yang-Mills theories.

Perhaps the main drawback of our study lies in the fact that the
new mass scale is not renormalization-group invariant; for example, we
can read off from Eq.\re{21} that
\begin{equation}
m_k^2=\frac{1}{16\pi^2}\, \Lambda^2 (1-\E^{2t}), \quad
t\equiv \ln \frac{k}{\Lambda}\leq0. \label{23}
\end{equation}
The new mass scale $m_k$ is necessarily proportional to $\Lambda$,
because there simply is no other scale in our system. But contrary to
the gauge coupling or the gauge parameter, which can be made
independent of $\Lambda$ by adjusting the bare parameters, the
$\Lambda$ dependence of $m_k$ persists, since there is no bare mass
parameter to adjust. One may speculate that this problem is solved by
``renormalization group improvement'' of the kind
\begin{equation}
\Lambda^2\to \Lambda^2\, \E^{-\frac{3\cdot 16\pi^2}{22 g^2(\Lambda)}},
\label{24}
\end{equation}
which upon insertion into Eq.\re{23} leads to a $\Lambda$-independent
mass scale for $k\to 0$. Obviously, our perturbative calculation can
never produce the RHS of Eq.\re{24}, but a nonperturbative study of
the renormalization group flow should result in such a structure (in a
different context, such a mechanism has been observed in 
\cite{Ellwanger:1998wv}).

Employing the measured values of the strong coupling constant at
various renormalization points, we can determine the order of
magnitude of the new mass scale: $m\equiv m_{k\to 0}= {\cal O}(1)$MeV,
e.g., $m\simeq5.74$MeV for $\alpha_{\text{s}}(M_{\text{Z}})=0.12$ or
$m\simeq0.68$MeV for $\alpha_{\text{s}}(10$GeV$)=0.18$ (the difference
between these numbers arises from the fact that the initial values for
the coupling are not related by a pure one-loop running). Of physical
interest are the masses of the solitonic excitation in this effective
theory. Unfortunately, there are no numerical results available for
theories with higher-derivative order, so that we have to resort to
results for an action identical to the first line of Eq.\re{21}. For
this model, the mass of the lowest lying states are approximately
given by $M\simeq{\cal O}(10^3) \sqrt{q}\, m$, where $q$ denotes the
value of the coefficient in front of the
$(\partial_\mu\mn\times\partial_\nu\mn)^2$ term
\cite{Gladikowski:1997mb,Battye:1999zn}. For couplings of order 1, we
end up with soliton masses of the order of $M\sim{\cal O}(1)$GeV; this
is in accordance with lattice results for glue ball masses: e.g., 
$M_{\text{GB}}\simeq 1.5$GeV for the lowest lying state in SU(2)
\cite{Teper:1998kw}. Of course, this rough and speculative estimate
should not be viewed as a ``serious prediction'' of our work.

With all these reservations in mind, the Faddeev-Niemi conjecture
about possible low-energy degrees of freedom of Yang-Mills theories
provides an interesting working hypothesis which deserves further
exploration. 

\section*{Acknowledgment}

The author wishes to thank W.~Dittrich for helpful conversations and
for carefully reading the manuscript. Furthermore, the author
profited from discussions with T.~Tok, K.~Langfeld and
A.~Sch\"afke. This work was supported in part by the Deutsche
Forschungsgemeinschaft under DFG GI 328/1-1.

\section*{Appendix}

\renewcommand{\thesection}{\mbox{\Alph{section}}}
\renewcommand{\theequation}{\mbox{\Alph{section}.\arabic{equation}}}
\setcounter{section}{0}
\setcounter{equation}{0}

\section{Differential operators, tensors, currents, etc.}

This appendix represents a collection of differential operators and
other tensorial quantities which are required in the main text. 

The Faddeev-Popov determinant $\Delta_{\text{FP}}$ in Eq.\re{7}
and\re{10} for covariant gauges involves the operator (in one-loop
approximation)
\begin{equation}
-\partial_\mu D_\mu(\mA)\bigl|_{C=0=\mW}
 =-\partial^2\mathbbm{1}_{\text{c}} 
 +(\partial^2\mn\otimes\mn -\mn\otimes\partial^2\mn)
 +(\partial_\mu\mn\otimes\mn-\mn\otimes\partial_\mu\mn)
 \partial_\mu, \label{A0}
\end{equation}
so that $\Delta_{\text{FP}}= \det \bigl(-\partial_\mu
D_\mu(\mA)\bigl|_{C=0=\mW} \bigr)$.

The objects occurring in the exponent of Eq.\re{7} are defined as
follows:
\begin{eqnarray}
&&M^C_{\mu\nu}:= -\partial^2 \delta_{\mu\nu} +\partial_\mu
\partial_\nu -\frac{1}{\alpha} \partial_\mu \partial_\nu
+\frac{1}{\alpha} \partial_\mu\mn \cdot \partial_\nu \mn\nonumber\\
&&M^{\mW}_{\mu\nu} := -\partial^2 \delta_{\mu\nu}
\mathbbm{1}_{\text{c}}+\partial_\mu \partial_\nu\mathbbm{1}_{\text{c}}
-\frac{1}{\alpha} \partial_\mu \partial_\nu\mathbbm{1}_{\text{c}}
 -\partial_\mu \mn \otimes \partial_\nu \mn + \partial_\nu \mn \otimes
 \partial_\mu \mn\nonumber\\
&&\mQ^{C}_{\mu\nu}:= \frac{1}{\alpha} \bigl( \partial_\mu\mn
\partial_\nu +\partial_\nu \mn \partial_\mu +\partial_\mu\partial_\nu
\mn\bigr)\nonumber\\
&&K^C_\mu:=\partial_\nu (\mn\cdot \partial_\nu \mn \times
\partial_\mu \mn) + \frac{1}{\alpha} \partial_\mu\mn\cdot
\partial^2\mn\times \mn \nonumber\\
&&\mK_\mu := \frac{1}{\alpha} \partial_\mu(\mn \times \partial^2
\mn).  \label{A1}
\end{eqnarray}
The determinants in Eq.\re{10} employ several composites of these
operators. Since we first perform the $C$ integration, the resulting
determinant involves only $M^C$, whereas the $\mW$ determinant also
receives contributions from the mixing term $\mQ^C$,
\begin{equation}
\overline{M}^{\mW}_{\mu\nu}={M}^{\mW}_{\mu\nu}+\widetilde{\mQ}^C_{\mu\kappa}
(M^C)^{-1}_{\kappa\lambda} \mQ^C_{\lambda\nu}. \label{A2}
\end{equation}
Here, $\widetilde{\mQ}$ is defined via partial integration
\begin{equation}
\int (\mQ^C_{\mu\nu} \mW_\nu) f_\nu\stackrel{\text{i.b.p}}{=} \int
\mW_\mu \widetilde{\mQ}^C_{\mu\nu} f_\nu, \label{A3}
\end{equation} 
and $f_\nu$ denotes an arbitrary test function. 

The last determinant in Eq.\re{10} arises from the $\mF$ integration
and receives contributions from the relevant parts of the exponent of
Eq.\re{9}, which we denote by
\begin{equation}
Q^{\mF}_\mu:= \I \bigl( -\partial_\mu \mathbbm{1}_{\text{c}} +
\partial_\mu\mn \otimes \mn\bigr), \label{A4}
\end{equation}
so that $\delta(\mX)\to \int {\cal D}\mF \exp (-\int\mW_\mu \cdot
Q^{\mF}_\mu \mF)$ to one-loop order.  Employing a notation similar to
Eq.\re{A3}, the differential operator accompanying the term $\sim
\mF\mF$ in the exponent finally reads $\widetilde{Q}^{\mF}_\mu
(\overline{M}^{\mW}_{\mu\nu})^{-1} Q^{\mF}_\nu$. Integrating the $\mF$
field along the imaginary axis leads to the last determinant in
Eq.\re{10}.

In Sect.~\ref{nfluc}, we employ the inverse gluon propagator
$(\Delta^{\text{YM}}[\mA^{\text{S}}])^{-1}$ coupled to all orders to
the soft $\mn$ field fluctuations. For an explicit representation, we
need the covariant derivative,
\begin{equation}
D_\mu[\mn]=\partial_\mu \mathbbm{1}_{\text{c}}+ 
\mn\otimes\partial_\mu\mn -\partial_\mu\mn\otimes\mn, \label{A5}
\end{equation}
where we have inserted the soft gauge potential Eq.\re{18} into the
covariant derivative. The inverse gluon propagator Eq.\re{20} then
reads
\begin{eqnarray}
(\Delta^{\text{YM}}[\mn])^{-1}_{\mu\nu}
&=&\mathbbm{1}_{\text{c}}\delta_{\mu\nu}
-2(\mn\otimes\partial_\lambda\mn -\partial_\mu\mn\otimes\mn)
  \partial_\lambda\delta_{\mu\nu}
+(\mn\otimes\partial_\nu\mn-\partial_\nu\mn\otimes\mn)\partial_\mu
\nonumber\\
&&+(\mn\otimes\partial_\mu\mn-\partial_\mu\mn\otimes\mn)\partial_\nu 
-(\mn\otimes\partial^2\mn-\partial^2\mn\otimes\mn)\delta_{\mu\nu}
+(\partial_\lambda\mn)^2\mn\otimes\mn \delta_{\mu\nu}\nonumber\\
&&+\partial_\lambda\mn\otimes\partial_\lambda\mn \delta_{\mu\nu}
-(2\partial_\mu\mn\otimes\partial_\nu\mn -\partial_\nu\mn\otimes
   \partial_\mu\mn)\nonumber\\
&&+(\mn\otimes\partial_{\mu\nu}\mn -\partial_{\mu\nu}\mn\otimes\mn)
-(\partial_\mu\mn\cdot\partial_\nu\mn)\mn\otimes\mn. \label{A6}
\end{eqnarray}

\section{Nonlocal terms}
During the Gaussian integration over the $C$, $\mF$ and $\mW$ fields
in Sect.~\ref{without}, several nonlocal terms arise from the
completion of the square in the exponent. Here, we shall give reasons
why they can be neglected. Let us exemplarily consider the simplest
nonlocal contribution arising from the $C$ integration:
\begin{equation}
K^C (M^C)^{-1} K^C= (\mn \cdot \partial_\lambda\mn\times
\partial_{\lambda\mu} \mn) \left(\frac{1}{-\partial^2
    +\partial\mn\cdot \partial\mn} \right)_{\mu\nu} (\mn\cdot
\partial_\kappa\mn\times \partial_{\kappa\nu} \mn). \label{B1}
\end{equation}
Within the calculation of the determinants, we expanded the inverse
operator assuming that $\partial\mn\cdot \partial\mn\ll
-\partial^2$. This was justified, since the derivative operator acts
on the test function space with momenta $p$ between $k$ and $\Lambda$,
which are large compared to the conjectured slow variation of the
$\mn$ field. 

In the present case, the situation is different, because the
derivative term $-\partial^2$ acts only on the $\mn$ field and its
derivatives to the right (there is no test function to act on). In
other words, the nonlocal terms are only numbers, not operators. The
derivatives can thus be approximated by the (inverse) scale of
variation of the $\mn$ field or its derivatives which is much smaller
than $k$ or $\Lambda$. This implies that the nonlocal terms do not
depend on $k$ or $\Lambda$, so that they cannot contribute to the flow
of the couplings. 

For example, a reasonable lowest-order approximation of the RHS of
Eq.\re{B1} is given by its local limit,
\begin{equation}
K^C (M^C)^{-1} K^C= (\mn \cdot \partial_\lambda\mn\times
\partial_{\lambda\mu} \mn) \left(\frac{1}{
    \partial\mn\cdot \partial\mn} \right)_{\mu\nu} (\mn\cdot
\partial_\kappa\mn\times \partial_{\kappa\nu} \mn)+\dots, \label{B2}
\end{equation}
where it is obvious that these terms do not contribute to the desired
Wilsonian effective action. The same line of argument holds for all
nonlocal terms appearing in Sects.~\ref{without} and \ref{nfluc}.

\section{Momentum integrals}
Several standard integrals appear in the integration over the momentum
shell $[k,\Lambda]$ in Sect.~\ref{without}. One basic formula is given
by
\begin{equation}
\int\limits_{[k,\Lambda]} 
\frac{d^4p}{(2\pi)^4} \, \frac{p_\lambda p_\kappa p_\mu
  p_\nu}{p^8} =\frac{1}{3} \frac{1}{64\pi^2} \ln \frac{\Lambda}{k}\,
\bigl( \delta_{\lambda\kappa}\delta_{\mu\nu} +\delta_{\lambda\mu}
\delta_{\kappa\nu} +\delta_{\lambda\nu}
\delta_{\kappa\mu}\bigr). \label{C1}
\end{equation}
From this formula, we can also deduce upon index contraction that
\begin{eqnarray}
\int\limits_{[k,\Lambda]} 
\frac{d^4p}{(2\pi)^4} \,\frac{p_\mu p_\nu}{p^6} = \frac{1}{32\pi^2}
\, \ln \frac{\Lambda}{k},\quad&&
\int\limits_{[k,\Lambda]} 
\frac{d^4p}{(2\pi)^4} \,\frac{1}{p^4} = \frac{1}{8\pi^2}
\, \ln \frac{\Lambda}{k}. \label{C3}
\end{eqnarray}
The last integral is, of course, standard and can be used to prove
Eq.\re{C1} in addition to symmetry arguments. The same philosophy
applies to the second type of integrals:
\begin{eqnarray}
\int\limits_{[k,\Lambda]} 
\frac{d^4p}{(2\pi)^4} \,\frac{p_\mu p_\nu}{p^4} = \frac{1}{64\pi^2}
\, (\Lambda^2-k^2)\, \delta_{\mu\nu}, \quad&&
\int\limits_{[k,\Lambda]} 
\frac{d^4p}{(2\pi)^4} \,\frac{1}{p^2} = \frac{1}{16\pi^2}
\, (\Lambda^2-k^2). \label{C5}
\end{eqnarray}

\end{document}